\documentclass[aps,11pt,prc,preprint,superscriptaddress,nofootinbib]{revtex4}

\usepackage[usenames]{color}
\usepackage{graphicx}
\usepackage{amsmath}
\usepackage{amsfonts}
\usepackage{amssymb}
\usepackage{mathrsfs}
\usepackage{bm}
\usepackage{verbatim}

\newcommand{\beq}{\begin{equation}}
\newcommand{\eeq}{\end{equation}}
\newcommand{\bea}{\begin{eqnarray}}
\newcommand{\eea}{\end{eqnarray}}

\newcommand{\mlo}{M_{\text{lo}}}
\newcommand{\mhi}{M_{\text{hi}}}
\newcommand{\chn}[3]{{{}^{#1}{#2}_{#3}}}
\newcommand{\cs}[2]{\chn{#1}{S}{#2}}

\newcommand{\y}{\text{Y}}

\preprint{CTP-SCU/2013002}
\preprint{JLAB-THY-13-1687}
\preprint{INT-PUB-13-024}

\begin{document}

\title{Improved convergence of chiral effective field theory for ${}^1S_0$ of $NN$ scattering}
\author{Bingwei Long}
\email{bingwei@scu.edu.cn}
\affiliation{Department of Physics, Sichuan University, 29 Wang-Jiang Road, Chengdu, Sichuan 610064, China}
\affiliation{Excited Baryon Analysis Center (EBAC), Jefferson Laboratory, 12000 Jefferson Avenue, Newport News, VA 23606, USA}

\date{\today}

\begin{abstract}
It is argued that the fine tuning due to the unnaturally large, generalized effective range in the ${}^1S_0$ channel of $NN$ scattering must be incorporated in order for one to obtain satisfactory convergence for chiral effective field theory. Without the proposition of perturbative one-pion exchange, an effective field theory with the spin-0, isospin-1 dibaryon is developed to account for this fine tuning, and is demonstrated up to $\mathcal{O}(Q^1)$ where the leading irreducible two-pion exchange arises. The approach shown in the paper results in rapid convergence of the ${}^1S_0$ partial-wave amplitude, though at the cost of an additional parameter at each order.
\end{abstract}

\maketitle

\section{Introduction\label{sec_intro}}

In the framework of chiral effective field theory (ChEFT), Weinberg's original power counting (WPC)~\cite{Weinberg:1990-1991, Ordonez:1993-1995, Epelbaum:1998ka-1999dj, Entem:2001cg-2002sf} for nucleon-nucleon scattering requires that at leading order (LO) in the $\cs{1}{0}$ channel a constant contact interaction and one-pion exchange (OPE) be fully iterated. However, a large discrepancy exists between the resulting EFT LO~\cite{Epelbaum:1998ka-1999dj, Nogga:2005hy, YangOhio} and partial-wave analysis (PWA) by, say, the Nijmegen group~\cite{Stoks:1993tb}, suggesting rather slow convergence of ChEFT expansion. One possibility is that there may exist an unexpected infrared mass scale due to fine tuning of quantum chromodynamics (QCD) such that momentum (or energy) dependence must be somehow incorporated into the LO short-range interaction, as opposed to WPC~\cite{Kaplan:1996nv}. Reference~\cite{Steele:1998zc} showed that such an infrared mass scale can be manifested by the inverse \emph{generalized} effective range in the modified effective range expansion (ERE) for $\cs{1}{0}$ (to be defined model-independently in the paper), $2/\widetilde{r} \sim 100$ MeV, whereas its natural value would have been around the breakdown scale of ChEFT: $\mhi \sim m_\sigma \simeq 600$ MeV with $m_\sigma$ being the mass of the $\sigma$ meson. This fine tuning requires resummation of $k\widetilde{r}/2$ to all orders, where $k$ is the magnitude of the center-of-mass (CM) momentum.

The large value of the $\cs{1}{0}$ scattering length, defined by the regular ERE near threshold, is yet another, albeit much better known, fine tuning of QCD~\cite{Beane:2006mx}. The two fine tunings do not seem to be correlated though, since $\widetilde{r}$ has more to do with the energy or momentum dependence of short-range forces while $a$ is closely related to the constant part. Originally designed to deal with the fine tuning of $a$, the machinery of Ref.~\cite{Kaplan:1996nv}, interestingly, facilitates resummation of $k\widetilde{r}/2$ to all orders: Introduce an auxiliary field, called a dibaryon field and denoted by $\bm{\phi}$, that has the same set of quantum numbers as the $\cs{1}{0}$ partial wave (baryon number 2, parity even, spin 0, and isospin 1) and the $s$-channel exchange of $\bm{\phi}$ will bring the desired energy dependence to the LO potential. It is my goal to show in this paper how this machinery can be generalized to include systematically higher-order corrections, in particular, those of irreducible two-pion exchanges (TPEs). This is part of our efforts~\cite{Long:2007vp, Long:2011qx-2011xw, Long:2012ve} to modify WPC and build a consistent and efficient power counting for chiral nuclear forces.

This goes beyond those works that modify WPC in order to respect renormalization group (RG) invariance~\cite{Barford:2002je, Birse:2005um, Nogga:2005hy, Valderrama:2009ei, Long:2012ve}, in which fine tuning of momentum-dependent $\cs{1}{0}$ counterterms were not particularly considered. However, the findings of Refs.~\cite{Barford:2002je, Birse:2005um, Long:2012ve} make the fine tuning of $\widetilde{r}$ appear less surprising than it would for WPC: In the natural chiral system RG invariance would require that the momentum dependence of $\cs{1}{0}$ counterterms arise as $\mathcal{O}(Q/\mhi)$ correction to LO, as opposed to the underestimation of $\mathcal{O}(Q^2/\mhi^2)$ by WPC.

Another line of investigation on applying the dibaryon fields to chiral nuclear forces can be found in Refs.~\cite{Soto:2007pg-2009xy-2011tb, Ando:2011aa}, in which the dibaryon fields are used for both $\cs{1}{0}$ and $\cs{3}{1}$ whereas only the spin-0 dibaryon is employed in my approach. In addition, the central Yukawa part of OPE [see Eq.~\eqref{eqn_defC0}] is treated as a perturbation in these works along the line of so-called KSW counting~\cite{Kaplan:1998tg_1998we}, whereas OPE is considered nonperturbative in this work [see the discussion below Eq.~\eqref{eqn_defC0}].

For the notation to be more compatible with the literature, I depart from the convention adopted in our previous papers~\cite{Long:2011qx-2011xw, Long:2012ve} and denote the order of EFT amplitudes by their absolute size rather than their relative size compared to LO. Therefore, nonperturbative LO will be labeled in the paper as $\mathcal{O}(Q^{-1})$, which is the scaling for any nonperturbative, nonrelativistic scattering amplitudes, $\mathcal{O}(Q/\mhi)$ corrections to LO as $\mathcal{O}(Q^0)$, and so on.

Note that the WPC LO of $\cs{1}{0}$ has actually another issue, which is not directly related to the aforementioned slow convergence: WPC fails to prescribe a quark-mass dependent counterterm at LO which is, however, required by RG invariance~\cite{Kaplan:1996xu, Beane:2001bc}.
As a by-product of the technique discussed in this paper to tackle the slow convergence, the quark-mass issue is solved altogether.

I review in Sec.~\ref{sec_lo} the theory without fine tuning and introduce the unnaturally large, generalized effective range. I then show in Sec.~\ref{sec_dibaryon} how this fine tuning can be incorporated by utilizing the dibaryon field, and then demonstrate the corresponding power counting up to $\mathcal{O}(Q^1)$ where the leading TPE needs to be accounted for. Finally a summary is offered in Sec.~\ref{sec_conclusion}.

\section{Issues at leading order\label{sec_lo}}

To motivate the employment of the dibaryon field, I briefly review the original theory that does not include it. Since most of the points to be shown here were already made in the literature, the main function of this section is to establish the notation.

The leading Lagrangian terms concerning the $\cs{1}{0}$ channel of $NN$ scattering are~\cite{Weinberg:1990-1991, Kaplan:1998tg_1998we, Fleming:1999ee}
\begin{equation}
\begin{split}
\mathcal{L}_{NN} &= \frac{1}{2} (\partial_\mu \bm{\pi})^2 - \frac{1}{2}m_\pi^2 \bm{\pi}^2 + N^\dagger \left(i\partial_0 + \frac{\nabla^2}{2m_N} \right) N - \frac{g_A}{2f_\pi} N^\dagger \tau_a \sigma_i (\partial_i\pi_a) N \\
&\quad - \widehat{C}_0 (N^T P_a N)^\dagger N^T P_a N - C_0^{qm} m_\pi^2 \left( \frac{1 - \bm{\pi}^2/4 f_\pi^2}{1 + \bm{\pi}^2/4 f_\pi^2}\right) (N^T P_a N)^\dagger N^T P_a N \\
&\quad  +  \frac{C_2}{8} \left[(N^T P_a N)^\dagger N^T P_a(\overleftarrow{\nabla} - \overrightarrow{\nabla})^2 N + \text{H.c.} \right] + \cdots \, ,
\end{split}
\label{eqn_philessLag}
\end{equation}
where $m_\pi = 138$ MeV, $g_A = 1.26$, $f_\pi = 92.4$ MeV, $m_N = 939$ MeV, and $P_a$ is the spin-isospin projector for the $\cs{1}{0}$ channel:
\begin{equation}
P_a = \frac{1}{\sqrt{8}} \tau_2 \tau_a \sigma_2 \, .
\end{equation} 
The quark-mass term proportional to $C_0^{qm}$ is written in the so-called stereographic coordinates for $\bm{\pi}$~\cite{Weinberg:1968}. Not only does it bring $m_\pi^2$ dependence to the contact interactions but it produces a nonderivative $\pi \pi NNNN$ vertex.

The leading $\cs{1}{0}$ amplitude by WPC is resummation of OPE and a constant $\cs{1}{0}$ counterterm to all orders:
\begin{equation}
V^{(-1)}(q) = V_\y(q) + C_0 \, ,
\end{equation} 
where the Yukawa potential and $C_0$ are defined as
\begin{equation}
V_\y(q) \equiv -\frac{4\pi}{m_N} \frac{\alpha_\pi m_\pi^2}{q^2 + m_\pi^2} \, , \quad C_0 \equiv \widehat{C}_0 + \frac{4\pi \alpha_\pi}{m_N} \, .
\label{eqn_defC0}
\end{equation} 
Here $\vec{q} \equiv \vec{p}\,' - \vec{p}$, with $\vec{p}\,'$ ($\vec{p}\,$) being the outgoing (incoming) momentum in the CM frame, and $\alpha_\pi^{-1} \equiv 16\pi f_\pi^2/g_A^2 m_N \sim 290 \text{MeV}$.
If $\alpha_\pi^{-1}$ is chosen to be an ultraviolet (UV) mass scale, the resulting power counting is KSW. This is especially plausible if the degrees of freedom of the delta-isobar are integrated out, and hence the delta-nucleon mass splitting $\simeq 300$ MeV becomes the breakdown scale. However, I choose to work in a more general scenario that keeps open the possibility of incorporating the delta-isobar, that is, $\alpha_\pi^{-1}$ is considered an infrared mass scale in the paper: $\alpha_\pi^{-1} \sim \mlo$. In addition, this allows for exploring situations where $m_\pi^2$ becomes so large that $m_\pi \alpha_\pi \sim 1$.

\begin{figure}
 \includegraphics[scale=0.4]{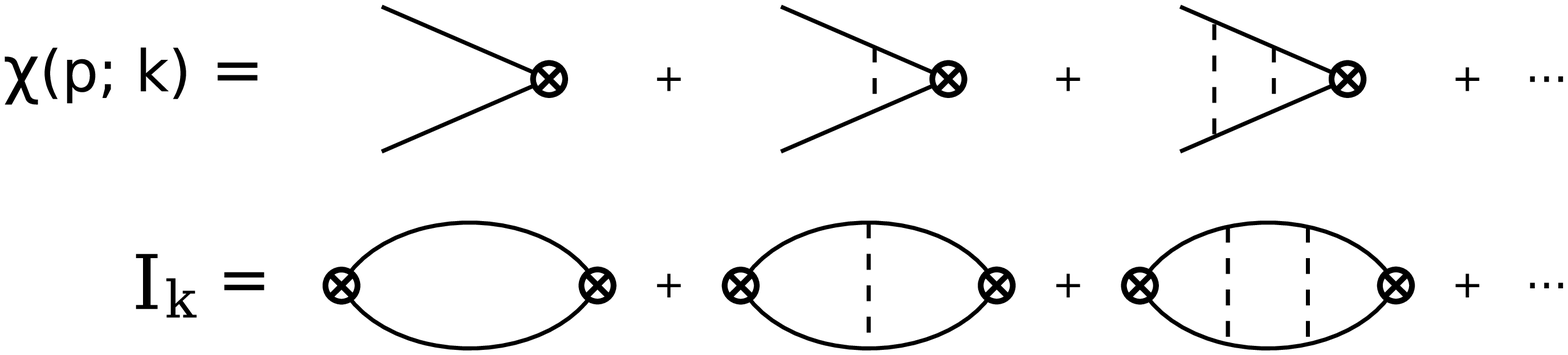}
\caption{Diagrammatic representation of $\chi(p; k)$ and $I_k$. Here the solid (dashed) lines represent the nucleon (pion) propagator, and the crossed circles represent no interactions.
\label{fig_chiIk}} 
\end{figure}

The LO $\cs{1}{0}$ amplitude can be written as~\cite{Kaplan:1996xu}
\begin{equation}
T^{(-1)}(\vec{p}\,', \vec{p}\,; k) = T_\y(\vec{p}\,', \vec{p}\,; k) + \frac{\chi(p'; k)\chi(p; k)}{1/C_0 - I_k} \, ,
\label{eqn_TLO_philess}
\end{equation} 
where $T_\y$ is the fully nonperturbative iteration of $V_\y$ and
\begin{align}
\chi(p; k) &= 1 + \int \frac{d^3l}{(2\pi)^3} \frac{T_\y(\vec{l}, \vec{p}\,; k)}{E - \frac{{l}^2}{m_N} + i\epsilon} \label{eqn_chiE} \, , \\
I_k &= \int \frac{d^3l}{(2\pi)^3} \frac{\chi(l; k)}{E - \frac{{l}^2}{m_N} + i\epsilon} \label{eqn_IE}  \, ,
\end{align}
with $E = k^2/m_N$ being the CM energy. Shown diagrammatically in Fig.~\ref{fig_chiIk}, the expansions of $\chi_k \equiv \chi(k; k)$ and $I_k$ in powers of $V_\y$ suggest that $\chi_k$ is finite while $I_k$ has divergences $\propto \frac{m_N}{4\pi} [\beta_0 \Lambda + \beta_1 \alpha_\pi m_\pi^2 \ln (\Lambda/\mu)]$, where $\Lambda$ is the UV momentum cutoff, $\mu$ is an infrared renormalization scale, and $\beta_{0, 1}$ are numerical factors depending on the form of the regulator.

The divergent $m_\pi^2$ dependence of $I_k$ immediately calls WPC into question: Chiral-invariant $1/C_0$ cannot absorb such a chiral-symmetry breaking divergence~\cite{Kaplan:1996xu, Beane:2001bc}. (This is not entirely an academic issue that concerns only extrapolating lattice QCD results to the physical point of $m_\pi$. As mentioned below Lagrangian~\eqref{eqn_philessLag}, the $m_\pi^2$ divergence is related through chiral symmetry to one of the nonderivative $\pi\pi NNNN$ vertices, which may have phenomenological impacts.) In addition to the obvious workaround that is KSW counting, it was proposed in Ref.~\cite{Beane:2001bc} that promoting to LO the operator proportional to $d_2 m_\pi^2$--- replacing $1/C_0$ with $1/(C_0 + d_2 m_\pi^2)$ in Eq.~\eqref{eqn_TLO_philess}--- will renormalize the $m_\pi^2$ divergence of $I_k$. This is quite a striking statement because it is not immediately clear how $1/(C_0 + d_2 m_\pi^2)$, as a fractional function of $m_\pi^2$, can absorb a divergence proportional to $m_\pi^2$. As we see, the dibaryon field can resolve this issue in a more transparent way by allowing its mass to be renormalized by iterations of $V_\y$.

Although important, the short-range structure related to quark masses does not change the form of the $\cs{1}{0}$ amplitude as a function of the CM momentum and, hence, does not help resolve the other issue at LO which is more phenomenologically urgent, namely, the slow convergence of EFT expansion in comparison to PWA. Thus, for simplicity in the qualitative discussion of the convergence issue, I allow  $C_0$ to have nontrivial $m_\pi^2$ dependence before we proceed to serious calculations. With this caveat, one can use $1/C_0$ to cancel both divergences of $I_k$ and writes on-shell $T^{(-1)}$ as
\begin{equation}
T^{(-1)} = T_Y + \frac{\chi^2_k}{1/C_0^R(\mu) - I_k^R(\mu)} \, ,
\end{equation}
where $I_k^R(\mu)$ is the $\mu$ dependent, finite part of $I_k$~\cite{Long:2012ve}. After fitting to the PWA, one immediately observes the slow convergence that is manifested by the rather large discrepancy between $T^{(-1)}$ and the PWA at low energies (see, e.g., Fig.~2 of Ref.~\cite{Long:2012ve}). For instance, the LO EFT predicts $\simeq 65^\circ$ for $k \simeq m_\pi$, whereas the Nijmegen PWA gives $\simeq 40^\circ$.

However, the next-to-leading-order (NLO) EFT curve fits the PWA well, which means that the NLO amplitude, $T^{(0)}$, is unexpectedly enhanced. Before any fine tuning is considered, the RG-invariant chiral theory for $\cs{1}{0}$ requires that (renormalized) contact coupling constants $C_0$ and $C_2$ scale as follows~\cite{Kaplan:1998tg_1998we, Barford:2002je, Birse:2005um, Long:2012ve}:
\begin{equation}
C_0^R \sim \frac{4\pi}{m_N} \frac{1}{\mlo}\, , \quad C_2^R \sim \frac{4\pi}{m_N} \frac{1}{\mlo^2 \mhi} \, , 
\label{eqn_C0C2philess}
\end{equation}
and hence the $\mathcal{O}(Q/\mhi)$ correction to LO \eqref{eqn_TLO_philess} is
\begin{equation}
T^{(0)} = \frac{C_2^R}{(C_0^R)^2} \frac{k^2\, {\chi_k}^2}{\left(1/C_0^R - I_k^R\right)^2} \, .
\label{eqn_oldT1}
\end{equation}
To see more easily how $C_0$ and $C_2$ are linked to the phase shifts, I recast $T^{(-1)} + T^{(0)}$ in the form of the modified ERE,
\begin{equation}
T^{(-1)} + T^{(0)} = T_\y - \frac{4\pi}{m_N} \, \frac{\chi_k^2}{-\frac{1}{\widetilde{a}(\mu)} + \frac{\widetilde{r}}{2} k^2 + \frac{4\pi}{m_N} I_k^R(\mu)}\left[1 + \mathcal{O}\left(\frac{k}{\mhi}\right) \right]\, ,
\label{eqn_dibaryonless_ERE}
\end{equation} 
where
\begin{align}
\frac{1}{\widetilde{a}(\mu)} &= \frac{4\pi}{m_N} \frac{1}{C_0^R} \sim \mlo \, , \\
\frac{\widetilde{r}}{2} &= -\frac{4\pi}{m_N} \frac{C_2^R}{(C_0^R)^2} \sim \frac{1}{\mhi} \, . \label{eqn_scale_r_philess}
\end{align} 
Unlike $\widetilde{a}(\mu)$, which depends on the renormalization scale, $\widetilde{r}$ is well defined and its value can be extracted from the phase shifts~\cite{Steele:1998zc}:
\begin{equation}
\frac{\widetilde{r}}{2} = 1.55\, \text{fm} = \frac{1}{127\,\text{MeV}} \, .
\end{equation} 
The fact that $2/\widetilde{r} \ll \mhi \simeq 600$ MeV signals that the data do not faithfully support the proposed scaling for $\widetilde{r}/2$ in Eq.~\eqref{eqn_scale_r_philess}--- the only avenue through which $\mhi$ could have suppressed $T^{(0)}$. To accommodate the empirical fact $\widetilde{r}/2 \sim 1/\mlo$, we need to develop a new scheme in which the ratios $k \widetilde{r}/2$ are resummed to all orders so that $\widetilde{r}k^2/2$ appears in the new LO amplitude rather than as a subleading correction.

The regular $\cs{1}{0}$ scattering length, defined as the zero-energy value of the amplitude, is also unnaturally large: $a_{\cs{1}{0}} \simeq -24$fm, compared with its would-be natural value of $\mathcal{O}(1)$ fm. It is tempting to consolidate the two fine tunings and to argue that they come from the same source, but ChEFT will not be able to do this because at $k = 0$ where $a_{\cs{1}{0}}$ is defined, $\widetilde{r}k^2/2$ does not contribute. In other words, as far as ChEFT is concerned, dialing $\widetilde{r}$ does not seem to have any effects on $a_{\cs{1}{0}}$.

Promoting nothing but $C_2$ to LO seems an obvious way to achieve the resummation of $\widetilde{r}k/2$. This was indeed proposed in Refs.~\cite{Kaplan:1996xu, Kaplan:1999qa, Beane:2001bc} and was numerically shown to work well for a range of cutoffs~\cite{Beane:2001bc}. By arranging some sophisticated runnings for $C_0$ and $C_2$, Ref.~\cite{Gegelia:2001ev} claimed that one can obtain analytically a renormalized LO amplitude. But Ref.~\cite{Phillips:1997xu} argued that iterating both $C_0$ and $C_2$ with different regularization schemes will lead to different results and that with a cutoff regulator the effective range allowed by the theory cannot be freely chosen--- the so-called Wigner bound~\cite{Wigner:1955zz, Phillips:1996ae}. So  it is still unclear whether one can promote $C_2$ alone without sacrificing RG invariance.

In addition to the aforementioned technical difficulty in renormalization, it would be quite surprising if fine tuning of $C_2$ can be isolated without contaminating operators with four or more derivatives. This is because multiple insertions of lower-order counterterms will generally renormalize higher-order ones through loops. Before fine tuning is considered, values of $C_{2n}$--- the coefficient of the $\cs{1}{0}$ four-nucleon operator with $2n$ derivatives--- are loosely correlated by $\mlo$ and $\mhi$, through dimensionless coefficients, $\theta_{2n}$, that are undetermined but are $\mathcal{O}(1)$~\cite{Long:2012ve}:
\begin{equation}
\frac{C_{2n}}{2} (p^{2n} + {p'}^{2n}) = \frac{4\pi}{m_N} \frac{\theta_{2n}}{\mlo^{n+1} \mhi^n} (p^{2n} + {p'}^{2n}) \, .
\label{eqn_C2nscaling}
\end{equation} 
(Note that Ref.~\cite{Barford:2002je} proposed different scalings for $C_{2n}$; hence, a different type of correlation ensues.) $\theta_2$ being tuned towards larger values while $\theta_0$ remains fixed effectively lowers $\mhi$, and such change of $\mhi$ propagates to $C_{2n}$ with $n \geqslant 2$.

Now I recklessly assume that the above correlation is still pertinent even when $C_{2n} (n \geqslant 1)$ are tuned to be so large that $\mhi$ is to be replaced by $\mlo$ in Eq.~\eqref{eqn_C2nscaling}. It then becomes apparent that all of $C_{2n}$ will be equally important. But an EFT with infinitely many unknown parameters at LO is not meaningful, unless we impose a stronger correlation among $\theta_{2n}$ such that $C_{2n}$ are determined by a finite number of LO parameters. Consider the following correlation of $C_{2n}$ by two parameters at LO, $\theta_0/\mlo$ and $\theta_2/\mlo$,
\begin{equation}
C_0 = \frac{4\pi}{m_N} \frac{\theta_0}{\mlo}\, , \quad
C_{2n} = \frac{4\pi}{m_N} \left(\frac{\theta_0}{\mlo}\right)^{n+1} \left(\frac{\theta_2}{\mlo}\right)^{n} \, .
\label{eqn_desiredC2n}
\end{equation} 
At tree level $p' = p = k \sim \mlo$, the sum of all $C_{2n} k^{2n}$ is
\begin{equation}
\sum_{n = 0}^{\infty} C_{2n} k^{2n} = \frac{4\pi}{m_N} \frac{1}{\mlo/\theta_0 - \theta_2 k^2/\mlo}\left[1 + \mathcal{O}\left(\frac{k^2}{\mhi\mlo}\right) \right]\, .
\label{eqn_sumC2n}
\end{equation}
The above summation resembles a tree-level $s$-channel exchange of $\bm{\phi}$ in $NN$ scattering:
\begin{equation}
\frac{\sigma y^2}{E + \Delta} = \frac{4\pi}{m_N} \frac{1}{\sigma\frac{4\pi}{m_N}\left( \frac{\Delta}{y^2} + \frac{k^2}{m_N y^2} \right)}\, ,
\label{eqn_Vphi_ped}
\end{equation}
with
\begin{equation}
\frac{4\pi \Delta}{m_N y^2} \sim \mlo\, , \quad \frac{4\pi}{m_N^2 y^2} \sim \frac{1}{\mlo} \, .
\end{equation}
Here $\sigma = \pm 1$, $\Delta$ is the mass splitting between $\bm{\phi}$ and two free nucleons, and $y$ is the $\phi N N$ coupling. As first shown in Ref.~\cite{Kaplan:1996nv} and to be reiterated in the next section, the $s$-channel $\bm{\phi}$ exchange will bring about the desired LO amplitude in the form of Eq.~\eqref{eqn_dibaryonless_ERE}, which is the ultimate justification for me to have chosen such a correlation as Eq.~\eqref{eqn_desiredC2n}.

\section{Power counting with the $\cs{1}{0}$ dibaryon field\label{sec_dibaryon}}

The first few Lagrangian terms involving $\bm{\phi}$ are~\cite{Kaplan:1996nv, Bedaque:1999vb}
\begin{equation}
\begin{split}
\mathcal{L}_{\phi} &= \sigma \bm{\phi}^\dagger \bm{\cdot} \left(i\mathscr{D}_0 + \frac{{\vec{\mathscr{D}}}^2}{4m_N} + \Delta \right) \bm{\phi} + y\left(\phi_a^\dagger N^T P_a N + \text{H.c.} \right) \\
&\quad + d_2\, m_\pi^2 \left(\frac{1 - \bm{\pi}^2/4 f_\pi^2}{1 + \bm{\pi}^2/4 f_\pi^2}\right) \bm{\phi}^\dagger\bm{\cdot\phi} + w_2 \, m_\pi^2 \left(\frac{1 - \bm{\pi}^2/4 f_\pi^2}{1 + \bm{\pi}^2/4 f_\pi^2}\right) \left(\phi_a^\dagger N^T P_a N + \text{H.c.} \right) \\
&\quad + d_4 \, m_\pi^4 \left(\frac{1 - \bm{\pi}^2/4 f_\pi^2}{1 + \bm{\pi}^2/4 f_\pi^2}\right)^2 \bm{\phi}^\dagger\bm{\cdot\phi} + \cdots \, ,
\end{split}
\label{eqn_phiLag}
\end{equation}
where $\mathscr{D}_\mu$ is the covariant derivative for an isovector field:
\begin{equation}
\mathscr{D}_\mu \bm{\phi} \equiv \partial_\mu \bm{\phi} - \left(1 + \frac{\bm{\pi}^2}{4 f_\pi^2}\right)^{-1} \left( \frac{\bm{\pi}}{f_\pi} \bm{\times}  \frac{\partial_\mu \bm{\pi}}{2 f_\pi} \right) \bm{\times \phi} \, .
\end{equation} 
Here I have normalized $\bm{\phi}$ so that $\sigma = \pm 1$. Later we see that fitting to the PWA results in $\sigma = -1$. The $d_2$ term are chiral-symmetry breaking and, as shown later, it needs to be at LO because of the fully iterated Yukawa potential. Terms that do not explicitly involve $\bm{\phi}$ but are needed in the paper are already shown in Lagrangian~\eqref{eqn_philessLag}.

\subsection{$\mathcal{O}(Q^{-1})$}

Following the argument that leads to Eqs.~\eqref{eqn_sumC2n} and \eqref{eqn_Vphi_ped}, I revise WPC so that the LO ``short-range'' potential is represented by an $s$-channel exchange of $\bm{\phi}$:
\begin{equation}
V^{(-1)} = V_\phi(E) + V_\y \, ,
\label{eqn_newV-1}
\end{equation}
where
\begin{equation}
V_\phi(E) \equiv \frac{\sigma y^2}{E + \Delta + d_2 m_\pi^2} \, .
\end{equation} 
Having both mass and kinetic terms of $\bm{\phi}$ at LO means that $V_\phi(E) \sim V_\y \sim \frac{4\pi}{m_N} \frac{m_\pi^2}{\mlo (m_\pi^2+Q^2)}$ and that renormalized $\Delta$, $d_2$, and $y$ scale as follows: 
\begin{equation}
\Delta^R \sim d_2^R m_\pi^2 \sim \frac{\mlo^2}{m_N} \quad \text{and}\quad (y^R)^2 \sim \frac{4\pi}{m_N} \frac{\mlo}{m_N}\, .
\label{eqn_Dyscaling}
\end{equation} 

The new LO potential is computationally equivalent to an energy-dependent $C_0$ in the dibaryon-less theory. With such an observation, we can write the new LO $\cs{1}{0}$ amplitude in an analogy to Eq.~\eqref{eqn_TLO_philess}:
\begin{equation}
\begin{split}
T^{(-1)} &= T_\y + \frac{\chi_k^2}{\sigma \frac{\Delta + d_2 m_\pi^2}{y^2} + \sigma \frac{k^2}{y^2 m_N} - I_k} \, .
\end{split}
\end{equation}
The necessity of having $d_2 m_\pi^2$ at LO is now clear; its assignment is to subtract the $m_\pi^2 \ln \Lambda$ divergence of $I_k$. It is worth stressing that promoting $d_2 m_\pi^2$ is independent of resumming the kinetic term of $\phi$, which is shown below to be responsible for generating $\widetilde{r}$ at LO. That is, even if we decide to live with the slow convergence of perturbative $\widetilde{r}/2$, the dibaryon still presents itself as a viable option for absorbing the $m_\pi^2 \ln \Lambda$ divergence of $I_k$. The term $d_2 m_\pi^2$ defying naive dimensional analysis has another consequence in addition to affecting the quark-mass dependence of the $\cs{1}{0}$ amplitude. As indicated by Lagrangian~\eqref{eqn_phiLag}, the promoted $d_2 m_\pi^2$ gives rise to an unsuppressed, nonderivative $\pi\pi\phi\phi$ coupling:
\begin{equation}
\mathcal{L}_{\pi \pi \phi \phi} = -d_2 m_\pi^2 \frac{\bm{\pi}^2}{2 f_\pi^2} \bm{\phi}^\dagger \bm{\cdot} \bm{\phi} \, . \label{eqn_pipiphiphi}
\end{equation} 

Again, renormalization of other Lagrangian parameters is perhaps most elucidated in the form of modified ERE:
\begin{equation}
T^{(-1)} = T_\y - \frac{4\pi}{m_N} \, \frac{\chi_k^2}{-\frac{1}{\widetilde{a}} + \frac{\widetilde{r}}{2} k^2 + \frac{4\pi}{m_N} I_k^R(\mu)} \, ,\label{eqn_newT0}
\end{equation}
with the generalized scattering length and generalized effective range defined for $\Lambda \to \infty$ as
\begin{align}
\frac{1}{\widetilde{a}(\mu)} &\equiv \frac{4\pi}{m_N}\left\{ \frac{\sigma\,\Delta^R}{(y^R)^2} + m_\pi^2 \frac{\sigma\,d_2^R }{(y^R)^{2}} \right\} \sim \mlo\, , \label{eqn_scaling_a}\\
\frac{\widetilde{r}}{2} &\equiv -\frac{4\pi}{m_N} \frac{\sigma}{m_N y^{2}} \sim \frac{1}{\mlo} \, ,\label{eqn_scaling_r}
\end{align}
where
\begin{equation}
\sigma\frac{\Delta^R}{(y^R)^2} \equiv \sigma\frac{\Delta}{y^2} - \beta_0 \Lambda \, , \quad 
\sigma\frac{d_2^R}{(y^R)^2} \equiv \sigma \frac{d_2 }{y^{2}} - \beta_1 \alpha_\pi \ln\left(\frac{\Lambda}{\mu}\right) \, , \quad \text{and} \quad
y^R \equiv y \, .
\end{equation}
Thus, we arrive at the desired scalings for $\widetilde{a}$ and $\widetilde{r}$. For a finite value of $\Lambda$, $\widetilde{r}$ has residual $\Lambda$ dependence that vanishes at the rate of $1/\Lambda$. The details of the numerical calculations are set up later, but I would like to remark that for $\Lambda = 800$ MeV, $\widetilde{r}/2$ is found to be $1/(115\text{MeV})$, which is consistent with its value stated in Ref.~\cite{Steele:1998zc}. Equation~\eqref{eqn_scaling_r} tells us that $\widetilde{r}/2$ and $\sigma$ must have opposite signs; therefore, $\sigma = -1$.

The form of the LO $\cs{1}{0}$ wave function in coordinate space is needed for later use. Since they are somewhat out of the main line of the physics, I relegate the technical details of its construction to Appendix~\ref{sec_wavefunciton}.

\subsection{$\mathcal{O}(Q^0)$}

The most general dibaryon Lagrangian is bound to have many redundant terms because there will not be enough observables to pin them down, due to the fact that $\bm{\phi}$ does not correspond to any particle appearing in asymptotic states. I choose to minimize the number of $\bm{\phi}$-related operators and to have four-nucleon contact operators be responsible for improving short-range interactions at subleading orders. This choice means that after the $\bm{\phi}$ exchange taking away the dominant part of short-range interactions, $C_0$ of the dibaryon Lagrangian represents higher-order effects: 
\begin{equation}
C_0 \sim \frac{4\pi}{m_N} \frac{1}{\mhi} \, .
\label{eqn_C0D2scaling}
\end{equation}
Note that I have slightly modified the scheme of Ref.~\cite{Kaplan:1996nv} in which $C_0$ was put on equal footing with the $\bm{\phi}$ exchange.

It is convenient to expand formally bare low-energy constants (LECs) to reflect the fact that even though the number of physical inputs must stay the same, their RG running may change at each order,
\begin{align}
\Delta^{B} &= \Delta^{(-1)} + \Delta^{(0)} + \Delta^{(1)} + \cdots \, , \\
d_2^B &= d_2^{(-1)}  + d_2^{(0)} + d_2^{(1)} + \cdots \, ,\\
y^B &= y^{(-1)} + y^{(0)} + y^{(1)} + \cdots \, ,\\
C_0^{B} &= C_0^{(0)} + C_0^{(1)} + \cdots \, , \label{eqn_C0B}\\
&\quad \cdots \nonumber
\end{align}
where the expansions are in powers of $1/\mhi$. For each parameter, the superscript of the leading term in its expansion marks the order it starts to contribute. For instance, since $C_0^{(0)}$ is the first term in Eq.~\eqref{eqn_C0B}, $\mathcal{O}(Q^0)$ will be the order $C_0$ occurs for the first time. However, in order to improve the readability of the manuscript, I make a few exceptions and drop the superscript $^{(-1)}$ for the first term of $\Delta$, $d_2$, and $y$.

NLO potential $V^{(0)}$ consists of only contact interactions and corrections to $\Delta$, $d_2$, and $y$:
\begin{equation}
V^{(0)} = C_0^{(0)} + 2 \left(y^{(0)} + w_2^{(0)} m_\pi^2\right) \frac{V_\phi}{y} - \sigma\left(\Delta^{(0)} + d_2^{(0)} m_\pi^2 + d_4^{(0)} m_\pi^4\right) \left(\frac{V_\phi}{y}\right)^2 \, .
\label{eqn_newV0}
\end{equation}
Despite the energy dependence of $V_\phi$, the technique shown in Appendix B of Ref.~\cite{Long:2012ve} is still useful for evaluating insertions of $V^{(0)}$. One can find a single insertion of $V^{(0)}$ to give rise to the generalized shape parameter, in addition to $m_\pi^4$ and $m_\pi^2$ corrections to $1/\widetilde{a}$ and $\widetilde{r}/2$, respectively:
\begin{equation}
\begin{split}
T^{(0)} = \frac{4\pi}{m_N} \frac{\left[ -(\frac{1}{\widetilde{a}})^{(0)} + \left(\frac{\widetilde{r}}{2}\right)^{(0)} k^2 + \widetilde{v}_2\, k^4\right] \, \chi_k^2}{\left(-\frac{1}{\widetilde{a}} + \frac{\widetilde{r}}{2} k^2 - \frac{4\pi}{m_N} I_k^R\right)^2} \, ,
\end{split}
\label{eqn_T1}
\end{equation}
where
\begin{align}
\left(\frac{1}{\widetilde{a}} \right)^{(0)} &= -m_\pi^4 \frac{4\pi}{m_N} \frac{\sigma}{y^2} \left(\sigma\frac{C_0^{(0)} d_2^2}{y^2} + 2\frac{w_2^{(0)} d_2}{y} - d_4^{(0)} \right) \, , \label{eqn_a0} \\
\left(\frac{\widetilde{r}}{2}\right)^{(0)} &= m_\pi^2 \frac{4\pi}{m_N} \frac{2\sigma }{y^3} \left(\frac{C_0^{(0)} d_2}{m_N y} + w_2^{(0)} \right) \, , \label{eqn_r0}\\
\widetilde{v}_2 &= \frac{m_N}{4\pi} C_0^{(0)} \frac{{\widetilde{r}}^2}{4} \, .
\label{eqn_v2}
\end{align}
Here I have chosen $\Delta^{(0)}$, $d_2^{(0)}$, and $y^{(0)}$ to be such that the chiral invariant parts of $1/\widetilde{a}$ and $\widetilde{r}/2$ and $m_\pi^2$ part of $1/\widetilde{a}$ retain their LO values:
\begin{align}
-\frac{\sigma \Delta}{y^2}\left( \sigma \frac{C_0^{(0)} \Delta}{y^2} + 2\frac{y^{(0)}}{y} - \frac{\Delta^{(0)}}{\Delta} \right) &= 0 \, , \\
-m_\pi^2 \frac{\sigma \Delta}{y^2} \left[ 2\sigma \frac{C_0^{(0)} d_2}{y^2} + 2 \left(\frac{w_2^{(0)}}{y} + \frac{y^{(0)}}{y} \frac{d_2}{\Delta} \right) - \frac{d_2^{(0)}}{\Delta} \right] &= 0\, ,\\
2 \frac{\sigma k^2}{m_N y^2} \left(\sigma \frac{C_0^{(0)} \Delta}{y^2} + \frac{y^{(0)}}{y} \right) &= 0\, .
\end{align}

$T^{(-1)} + T^{(0)}$ can be rewritten in the form of modified ERE:
\begin{equation}
T^{(-1)} + T^{(0)} = T_\y - \frac{4\pi}{m_N} \, \frac{\chi_k^2}{-\frac{1}{\widetilde{a}} + \frac{\widetilde{r}}{2} k^2 + \widetilde{v}_{2} k^{4} + \frac{4\pi}{m_N} I_k^R} + \mathcal{O}\left(\frac{Q^2}{\mhi^2} T^{(-1)}\right) \, .
\end{equation}
Power counting~\eqref{eqn_C0D2scaling} is then equivalent to estimating $\widetilde{v}_2$ as
\begin{equation}
\widetilde{v}_2 \sim \frac{1}{\mhi} \frac{\widetilde{r}^2}{4} \sim \frac{1}{\mlo^2 \mhi} \, ,
\label{eqn_scaling_v2}
\end{equation}
which is compatible with the value extracted in Ref.~\cite{Steele:1998zc},
\begin{equation}
\widetilde{v}_2 = \frac{\widetilde{r}^2/4}{550 \text{MeV}} \, , \quad \text{with} \quad \frac{\widetilde{r}}{2} = \frac{1}{127 \text{MeV}} \, ,
\end{equation}
and the value by this work for $\Lambda = 800$ MeV,
\begin{equation}
\widetilde{v}_2 = \frac{\widetilde{r}^2/4}{693 \text{MeV}} \, , \quad \text{with} \quad \frac{\widetilde{r}}{2} = \frac{1}{115 \text{MeV}} \, . 
\end{equation}

\subsection{$\mathcal{O}(Q^1)$}

\subsubsection{Residual counterterms}
It is, if only academically, interesting to ask how counterterms will scale if TPEs and higher-order multiple-pion exchanges are completely turned off while the strength of OPE remains unchanged, which can be achieved by taking $1/f_\pi^2 \to 0$ but keeping $m_N/f_\pi^2$ fixed. Higher-order counterterms in this scenario, referred to as ``residual counterterms'' in Ref.~\cite{Long:2012ve}, are responsible for all the subleading corrections, and the modified ERE is expected to be valid to all orders and to acquire $k^{2n}$ terms beyond $\widetilde{v}_2 k^4$,
\begin{equation}
T_\y - \frac{4\pi}{m_N} \, \frac{\chi_k^2}{-\frac{1}{\widetilde{a}} + \frac{\widetilde{r}}{2} k^2 + \sum\limits_{n = 2} \widetilde{v}_{n} k^{2n} + \frac{4\pi}{m_N} I_k^R}\, .
\end{equation} 
I wish to find out how $\widetilde{v}_n$ (for $n \geqslant 3$) scales in such a hypothetical scenario.

Next-to-next-to-leading-order (NNLO) amplitude $T^{(1)}$ includes two insertions of $C_0$, which can be compared with a single insertion of $C_2 (p^2 + {p'}^2)/2$:
\begin{equation}
\begin{split}
T^{(1)}_{2V^{(0)}+C_2^{(1)}}
&= \frac{4\pi}{m_N} \frac{\chi_k^2}{\left(-\frac{1}{\widetilde{a}} + \frac{\widetilde{r}}{2} k^2 + \frac{4\pi}{m_N} I_k^R\right)^2 }\left( \widetilde{v}_3 k^6 - \frac{\widetilde{v}^2_2 k^8}{-\frac{1}{\widetilde{a}} + \frac{\widetilde{r}}{2} k^2 + \frac{4\pi}{m_N} I_k^R} + \cdots \right) \, ,
\label{eqn_T2V1C2}
\end{split}
\end{equation} 
with
\begin{equation}
\widetilde{v}_3 \equiv \frac{\widetilde{v}_2^2}{\widetilde{r}/2} + \frac{m_N}{4\pi}\,C_2^{(1)} \frac{\widetilde{r}^2}{4} \, ,\label{eqn_v3}
\end{equation} 
where $\cdots$ refers to $m_\pi^2$ corrections to $1/\widetilde{a}$, $\widetilde{r}/2$, and $\widetilde{v}_2$, which are not pertinent to the present discussion. $C_2^{(1)}$ is not running with $\Lambda$, but it is nonetheless renormalized by a term quadratic in $C_0^{(0)}$ [$C_0^{(0)}$ is related to $\widetilde{v}_2$ through Eq.~\eqref{eqn_v2}]. With fine tuning having been accounted for by the resummation at LO, it is reasonable to expect naturalness to retain its power in counting. Therefore, the two terms contributing to $\widetilde{v}_3$ must have similar sizes, resulting in
\begin{equation}
\widetilde{v}_{3} \sim \frac{1}{\mlo^{3} \mhi^{2}} \quad\text{and}\quad C_{2} \sim \frac{4\pi}{m_N} \frac{1}{\mlo\mhi^{2}}\, .
\label{eqn_C2res}
\end{equation} 

More generally, one can show that $\widetilde{v}_{n+1}$ will have contributions, among others, from $n$ insertions of $C_0$ and one insertion of $C_{2n-2}(p^{2n} + {p'}^{2n})/2$:
\begin{equation}
\widetilde{v}_{n+1} = \frac{\widetilde{v}_2^{n}}{\widetilde{r}/2} + \frac{m_N}{4\pi}\,C_{2n-2} \frac{\widetilde{r}^2}{4} + \cdots\, ,
\end{equation} 
which leads to 
\begin{equation}
\widetilde{v}_{n+1} \sim \frac{1}{\mlo^{n+1} \mhi^{n}} \, ,
\quad
C_{2n-2} \sim \frac{4\pi}{m_N} \frac{1}{\mlo^{n-1}\mhi^{n}}\, .
\label{eqn_newC2nscaling}
\end{equation}

\subsubsection{Two-pion exchange}

Let us turn to the leading TPE, $V_{2\pi}$. Throughout our efforts~\cite{Long:2007vp, Long:2011qx-2011xw, Long:2012ve} to modify WPC, we have taken the position that the standard chiral counting does not need to change for pion-exchange diagrams~\cite{Weinberg:1990-1991}, which essentially describe long-range physics through nonanalytic functions of momenta. This means that since $V_{2\pi}$ is suppressed by $\mathcal{O}[Q^2/(4\pi f_\pi)^2]$ relative to OPE, the single insertion of $V_{2\pi}$ into the nonperturbative LO amplitude is also suppressed by $\mathcal{O}[Q^2/(4\pi f_\pi)^2]$.\footnote{However, I would like to remind the reader that a different point of view towards numerical factors of $\pi$ in chiral counting can be found in Ref.~\cite{Baru:2012iv}.}

But one still needs to determine what counterterms are required to renormalize the UV part of the single insertion of $V_{2\pi}$. They are called ``distorted-wave counterterms'' in Ref.~\cite{Long:2012ve} because a single insertion of $V_{2\pi}$ is equivalent to the matrix element of $V_{2\pi}$ between the LO wave functions--- the distorted wave for the LO potential~\cite{Valderrama:2009ei, Long:2012ve}---
\begin{equation}
\langle \psi_k | V_{2\pi} | \psi_k \rangle = 4\pi \int_{\sim \Lambda^{-1}} dr\, r^2 \psi^2_k(r)\, \widetilde{V}_{2\pi}^{(0)}(r) \, . 
\label{eqn_V2pimatelem}
\end{equation}
Near the origin, $\psi_k(r)$ can be expanded in powers of $(kr)^2$ [see Eq.~\eqref{eqn_psik_smallr}] and  $V_{2\pi} \propto 1/r^5$; therefore, one can find the superficial divergence of $\langle \psi_k | V_{2\pi} | \psi_k \rangle$ to be
\begin{equation}
\begin{split}
& 4\pi \int_{\sim \Lambda^{-1}} dr\, r^2 \psi^2_k(r)\, \widetilde{V}_{2\pi}^{(0)}(r) \\
&\; \propto \left( \frac{\chi_k}{V_\phi^{-1} - I_k} \right)^2 \left( \rho_0 \Lambda^4 + \rho_1 \widetilde{r} k^2 \Lambda^3 + \rho_2 \widetilde{r}^2 k^4 \Lambda^2 + \rho_3 \widetilde{r}^2 k^6 \ln \Lambda \right) + \text{F.T.} \, ,
\label{eqn_V2pisuper}
\end{split}
\end{equation} 
where ``F.T.'' refers to finite terms and $\rho_n$ have at most logarithmic dependence on $\Lambda$.

To identify the needed counterterms, we first notice that $C_2(p^2 + {p'}^2)/2$ produces $k^6 \chi_k^2/(V_\phi^{-1} - I_k)^2$, as suggested by Eqs.~\eqref{eqn_T2V1C2} and \eqref{eqn_v3}. Furthermore, one can show that other divergences with lower powers of $k^2$ than $k^6$ can be subtracted by corrections to $\Delta$, $y$, and $C_0$. With every piece put together, $V^{(1)}$ has the form:
\begin{equation}
V^{(1)} =  V_{2\pi} + \frac{C_2^{(0)}}{2}(p^2 + {p'}^2) + \lambda_0 + \lambda_{1}V_\phi(E) + \lambda_{2} V_\phi^2(E) + \lambda_{3} V^{3}_\phi(E) \, .\label{eqn_newV1}
\end{equation} 
It would be an unnecessary bore to write the expressions of $\lambda_{i}$ in terms of $\Delta^{(1)}$,  $y^{(1)}$, etc., because in practice all we need to know is that $\lambda_{i}$ are independent of energy or momenta. I have also stopped pursuing the complete understanding of $m_\pi^2$ dependence, which will be dealt with in a future publication.

It simplifies tremendously the work of establishing power counting that $V_{2\pi}$ does not demand for renormalization purposes more counterterms than the residual counting~\eqref{eqn_newC2nscaling} provides. I extrapolate this observation to any irreducible multiple-pion exchanges and conclude that all of $\cs{1}{0}$ counterterms are prescribed by power counting~\eqref{eqn_newC2nscaling}.

\subsection{Results}

Although the formal expressions shown earlier in this paper reveal the renormalization and the analytic structure of the amplitude at each order, a complete analytical calculation is still unlikely,  for quantities like $\chi(p; k)$ and $I_k$ cannot be computed analytically. Below I elaborate the setup of numerical calculations for the $\cs{1}{0}$ phase shifts and discuss the results.

The potentials are regularized with a separable momentum-space regulator:
\begin{equation}
V_\Lambda(\vec{p}\,', \vec{p}; E) = \exp \left(-\frac{{p'}^4 + p^4}{\Lambda^4}\right) V(\vec{p}\,', \vec{p}; E) \, .
\end{equation} 
Whereas the LO amplitude is generated nonperturbatively by solving the Lippmann-Schwinger equation for the LO potential~\eqref{eqn_newV0},
\begin{equation}
T^{(-1)} = V^{(-1)} + V^{(-1)}\,G\, T^{(-1)}\, ,
\end{equation}
where $G$ is the Schr\"odinger propagator, the subleading corrections are calculated through perturbative insertions of $V^{(0)}$ [Eq.~\eqref{eqn_newV0}] and $V^{(1)}$ [Eq.~\eqref{eqn_newV1}],
\begin{align}
T^{(0)} &= \left(1 + T^{(-1)}\, G\right) V^{(0)} \left(G\, T^{(-1)} + 1\right)\, , \\
T^{(1)} &= \left(1 + T^{(-1)}\, G\right) \left[V^{(1)} + V^{(0)}\left(G + G\, T^{(-1)}\,G\right)V^{(0)} \right] \left(G\, T^{(-1)} + 1\right)\, ,
\end{align}
in which I adopt from Ref.~\cite{Epelbaum:1998ka-1999dj} the delta-less version for $V_{2\pi}$.

\begin{figure}
\includegraphics[scale=0.7, clip=true]{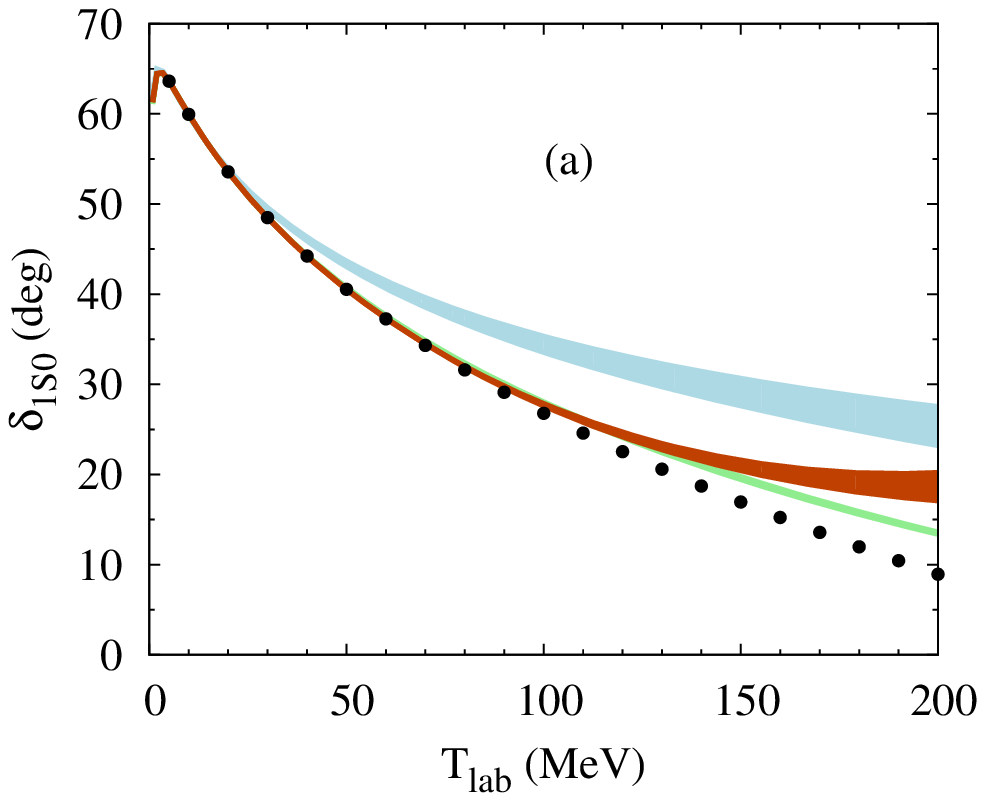} 
\includegraphics[scale=0.7, clip=true]{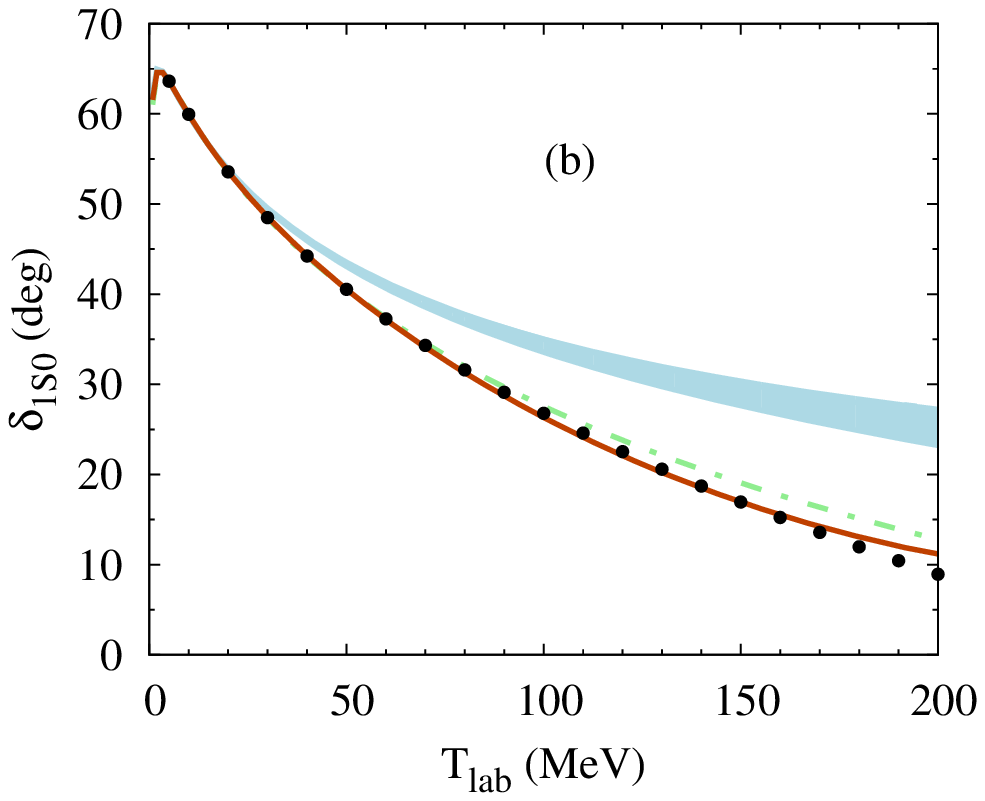}
\caption{\label{fig_tlab1s0}(Color online) $\cs{1}{0}$ phase shifts as a function of laboratory energy. The black dots are from the Nijmegen PWA~\cite{Stoks:1993tb}. (a) The light-blue, light-green, and dark-orange bands are $\mathcal{O}(Q^{-1})$, $\mathcal{O}(Q^0)$, and $\mathcal{O}(Q^1)$ calculated with $\Lambda = 0.6 - 2$ GeV. (b) $\mathcal{O}(Q^{0})$ (light-green dot-dashed line) and $\mathcal{O}(Q^1)$ (dark-orange solid line)  are plotted with $\Lambda = 1$ GeV.}
\end{figure}

Figure~\ref{fig_tlab1s0} shows the EFT results for $\cs{1}{0}$ phase shifts up to and including $\mathcal{O}(Q^1)$. In Fig.~\ref{fig_tlab1s0}(a), the LO is obtained by fitting to the PWA points at $T_\text{lab} = 5$ and $15$ MeV. At $\mathcal{O}(Q^0)$ and $\mathcal{O}(Q^1)$, 25 and 50 MeV are added, respectively. The bands are generated by $\Lambda = 0.6 - 2$ GeV. Above $\Lambda = 2$ GeV, the cutoff variation is smaller than one tenth degree at, e.g., $T_\text{lab} = 130$ MeV. Compared with the dibaryon-less theory~\cite{Valderrama:2009ei, Long:2012ve}, the new formulation fits much better to the PWA and converges rapidly, at the cost of one more short-range parameter at each order. The breakdown of convergence around $T_\text{lab} \simeq 200$ MeV is not surprising, for the delta-isobar is not explicitly considered.

The small correction provided by the leading TPE reassures its perturbative nature. This is in contrast to the WPC-based study of Ref.~\cite{Shukla:2008sp}, which suggested that in order to have perturbative multiple-pion exchanges the cutoff needs to be soft, and the speculation of Ref.~\cite{Birse:2010jr} that the delta-isobar may be the source of the slow convergence.

To get an idea of how much the fit can be further improved, I fit to PWA points at higher energies, and the results are shown in Fig.~\ref{fig_tlab1s0}(b). There the fitted PWA points are at $T_\text{lab} = 50$ MeV for $\mathcal{O}(Q^0)$ and $T_\text{lab} = 150$ MeV for $\mathcal{O}(Q^1)$, while the inputs for LO did not change from Fig.~\ref{fig_tlab1s0}(a). Since the cutoff dependence is no longer a concern, only $\Lambda = 1$ GeV is used for $\mathcal{O}(Q^0)$ and $\mathcal{O}(Q^1)$.

\section{Summary\label{sec_conclusion}}

I have considered the EFT expansion for $\cs{1}{0}$ of $NN$ scattering in which the generalized effective range  $\widetilde{r}/2$ is counted as an infrared length scale due to fine tuning of the underlying theory. The new expansion was made possible by an auxiliary, dibaryon field $\bm{\phi}$ that has the same quantum numbers as the $\cs{1}{0}$ partial wave~\cite{Kaplan:1996xu}. At LO, the fine tuning in question is incorporated by iterating the $s$-channel exchange of $\bm{\phi}$ to all orders. The price to pay for the nonperturbative treatment of $\widetilde{r}/2$ is an additional short-range parameter at each order, compared with the power counting for perturbative $\widetilde{r}/2$~\cite{Long:2012ve, Birse:2005um}.

I chose to minimize the number of $\bm{\phi}$-dependent operators and to use four-nucleon counterterms to account for subleading short-range forces. If irreducible multiple-pion exchanges were hypothetically turned off, power counting of the so-called residual counterterms could be considered. $C_{2n}$ would appear in the $\mathcal{O}(Q^{n+1}/\mhi^{n+1})$ corrections to LO and would scale as
\begin{equation}
C_{2n} \sim \frac{4\pi}{m_N} \frac{1}{\mlo^{n}\mhi^{n+1}}\, .
\end{equation} 
This counting was actually found to provide enough counterterms to absorb the divergences of TPEs, when they are turned back on and are inserted to the LO amplitude. Therefore, the above power counting is the final answer we were looking for. The numerical results showed much improved convergence of the EFT $\cs{1}{0}$ phase shifts.

In addition to resumming $\widetilde{r}/2$, the dibaryon field provides a transparent mechanism to deal with quark-mass-dependent contact operators that concern the $\cs{1}{0}$ channel. At LO the dibaryon field absorbs the logarithmic $m_\pi^2$ divergence by allowing its mass to be renormalized. Through chiral symmetry, this immediately calls for renormalization-driven promotion of the quark-mass dependent, nonderivative $\pi\pi \phi \phi$ coupling [see Eq.~\eqref{eqn_pipiphiphi}]. I also showed the $m_\pi^2$ dependence of the $\cs{1}{0}$ operators up to $\mathcal{O}(Q^0)$. A more complete study on the quark-mass dependence of low-energy $\cs{1}{0}$ scattering is reserved for a future publication.

\acknowledgments I thank Manuel Pavon Valderrama, Daniel Phillips, and Bira van Kolck for useful discussions. I am grateful for hospitality to the Institute for High Energy Physics in Beijing, the Institute for Modern Physics in Lanzhou, the National Institute for Nuclear Theory (INT) at the University of Washington, where part of the work was done, and the organizers of the INT program ``Light Nuclei from First Principle'' for making my participation of the program possible. This work is partly supported by the US DOE under contract No.DE-AC05-06OR23177 and is coauthored by Jefferson Science Associates, LLC under U.S. DOE Contract No. DE-AC05-06OR23177.

\appendix

\section{LO wave function\label{sec_wavefunciton}}

I follow the technique developed in Ref.~\cite{Kaplan:1996xu} to obtain the short-distance behavior of the LO $\cs{1}{0}$ wave function. In the limit $\Lambda \to \infty$, the LO potential has the following formal coordinate-space form:
\begin{equation}
\widetilde{V}^{(0)}(\vec{r}\,) = V_\phi(E)\delta^{(3)}(\vec{r}\,) + \widetilde{V}_\y(r) \, ,
\end{equation} 
where $\widetilde{V}_\y(\vec{r}\,)$ is the Fourier transform of the Yukawa potential and $V_\phi$ is defined in Eq.~\eqref{eqn_newV0}. The in-state, $S$-wave wave function formally satisfies
\begin{equation}
\left[-\frac{1}{m_N} \left(\frac{d^2}{dr^2} + \frac{2}{r} \frac{d}{dr} \right) + \widetilde{V}_\y - E\right] \psi_k(r) = - V_\phi(E) \psi_k(0) \delta^{(3)}(\vec{r}\,) \, ,
\label{eqn_schrodinger}
\end{equation} 
and can be written as a linear combination of the regular and irregular solutions to the Schr\"odinger equation for $\widetilde{V}_\y$,
\begin{equation}
\psi_k(r) = a(k)\mathcal{J}_k(r) + b(k)\mathcal{H}_k(r) \, ,
\label{eqn_psi_comb}
\end{equation} 
where $\mathcal{J}_k(r)$ and $\mathcal{H}_k(r)$ are normalized so that they satisfy
\begin{align}
\left[-\frac{1}{m_N} \left(\frac{d^2}{dr^2} + \frac{2}{r} \frac{d}{dr} \right) + \widetilde{V}_\y - E\right] \mathcal{J}_k(r) &= 0\, ,
\label{eqn_Jk} \\
\left[-\frac{1}{m_N} \left(\frac{d^2}{dr^2} + \frac{2}{r} \frac{d}{dr} \right) + \widetilde{V}_\y - E\right] \mathcal{H}_k(r) &= \delta^{(3)}(\vec{r}\,) \, , \label{eqn_Hk}
\end{align}
While $ \mathcal{J}_k(r) \to j_0(kr)$ for $r \to 0$, where $j_0(x)$ is the zeroth spherical Bessel function, $\mathcal{H}_k$ has the following form for  $r \to 0$:
\begin{equation}
\begin{split}
\mathcal{H}_k(r) = \frac{m_N}{4\pi} \left[ \frac{1}{r} \mathcal{B}(kr, \kappa_\pi r) - 2\kappa_\pi \mathcal{A}(kr, \kappa_\pi r) \ln \left(\mu r\right) \right] \, ,
\end{split}\label{eqn_coulomb}
\end{equation}
where $\kappa_\pi = m_\pi^2 \alpha_\pi$ and $\mathcal{A}(x, y)$ and $\mathcal{B}(x, y)$ are dimensionless functions that are analytic at $x, y = 0$. Using the above expression one can obtain the expansion of $\mathcal{H}_k(r)$ in powers of $(kr)^2$ and/or $(\kappa_\pi r)^2$ near the origin. 

For any cutoff regulator, the delta potential gets smeared away from the origin, up to a distance characterized by $\mathcal{R} \equiv \Lambda^{-1}$. Solution~\eqref{eqn_psi_comb} in fact governs the ``outside region'', $r \gtrsim \mathcal{R}$. Reference~\cite{Kaplan:1996xu} showed that the singularity of $\mathcal{H}_k(\mathcal{R})$ for $\mathcal{R} \to 0$ can be related to the divergences of $I_k$. To see this, notice that Eqs.~\eqref{eqn_Jk} and \eqref{eqn_Hk} indicate that a certain linear combination of $\mathcal{H}_k(r)$ and $\mathcal{J}_k(r)$ makes up the $S$-wave interacting Green function for the Yukawa potential: $G_Y(r; E) \equiv \langle r, Y_0^0(\theta, \phi) | \left(E - H_0 - V_\y + i\epsilon \right)^{-1} | \vec{x} = 0 \rangle $ with $H_0$ being the free two-nucleon Hamiltonian and $Y_0^0(\theta, \phi)$ the $S$-wave spherical harmonic. The divergence of $G_Y(0; E)$ is completely described by $\mathcal{H}_k(\mathcal{R} \to 0)$ and does not depend on the $\mathcal{J}_k$ part, since $\mathcal{J}_k(r)$ behaves well near $r = 0$. On the other hand, $I_k$ is precisely $G_Y(0; E)$, most easily seen from its diagrammatic representation in Fig.~\ref{fig_chiIk}. Now we can identify the divergences of $I_k$, $-\beta_0 \Lambda - \beta_1 \kappa_\pi \ln(\Lambda/\mu)$, with $\mathcal{H}_k(\mathcal{R} \to 0)$. This means that following subtractions in Eqs.~\eqref{eqn_scaling_a} and \eqref{eqn_scaling_r}, one can also use $V_\phi^{-1}$ to subtract the singularity of $\mathcal{H}_k(\mathcal{R} \to 0)$:
\begin{equation}
V_\phi^{-1} + \mathcal{H}_k(0) = \frac{m_N}{4\pi} \left(-\frac{1}{\widetilde{a}} + \frac{\widetilde{r}}{2} k^2\right) \, .
\label{eqn_VphiplusHk}
\end{equation} 

Substituting Eq.~\eqref{eqn_psi_comb} in Eq.~\eqref{eqn_schrodinger} and applying Eqs.~\eqref{eqn_Jk} and \eqref{eqn_Hk} on the left-hand side of Eq.~\eqref{eqn_schrodinger}, we find
\begin{equation}
b = - V_\phi(E) \psi_k(0) = b\left[\frac{a}{b} + \mathcal{H}_k(0) \right]\, .\label{eqn_bdef}
\end{equation} 
However, $\psi_k(0)$ is generally related to the LO off-shell $T$-matrix by
\begin{equation}
\psi_k(0) = 1 + \int \frac{d^3l}{(2\pi)^3} \frac{T^{(-1)}(\vec{l},\vec{k}; k)}{E - \frac{l^2}{m_N} + i\epsilon} \, .
\end{equation} 
Using Eq.~\eqref{eqn_TLO_philess}, with $C_0$ being replaced with $V_\phi(E)$, in the above equation, one finds
\begin{equation}
\psi_k(0) = \frac{\chi_k V_\phi^{-1}}{V_\phi^{-1} - I_k} = \frac{4\pi}{m_N} \frac{\chi_k V_\phi^{-1}}{-\frac{1}{\widetilde{a}} + \frac{\widetilde{r}}{2} k^2 - \frac{4\pi}{m_N} I_k^R} \, .
\label{eqn_psik0}
\end{equation}
Using the above expression and Eq.~\eqref{eqn_VphiplusHk} in Eq.~\eqref{eqn_bdef} gives
\begin{align}
b &= -\frac{\chi_k}{V_\phi^{-1} - I_k} \, , \\
\frac{a}{b} &= -\left[V_\phi^{-1} + \mathcal{H}_k(0)\right] = - \frac{m_N}{4\pi} \left(-\frac{1}{\widetilde{a}} + \frac{\widetilde{r}}{2} k^2 \right) \, .
\end{align}
It is worth noting the difference between this LO wave function and that of the dibaryon-less theory: While $b$ stays the same, $a/b$ now has $k^2$ dependence, in contrast to being a constant in the dibaryon-less theory~\cite{Kaplan:1996nv, Long:2012ve}, a consequence stemming from the fact that the LO contact interaction now provides two inputs for the outside wave function.

We can now put these back into Eq.~\eqref{eqn_psi_comb} to have a more clear expression of the short-distance behavior of $\psi_k(r)$:
\begin{equation}
\psi_k(r) = -\frac{\chi_k}{V_\phi^{-1} - I_k} \frac{m_N}{4\pi} \left\{\frac{1}{r} \mathcal{A}(kr, \kappa_\pi r) - \left[2 \kappa_\pi \ln(\mu r) - \frac{1}{\widetilde{a}} + \frac{\widetilde{r}}{2} k^2\right] \mathcal{B}(kr, \kappa_\pi r) \right\} \, .
\label{eqn_psik_smallr}
\end{equation} 
This expression is in principle accurate only for $\mathcal{R} \ll r \ll k^{-1}$, and the specification of the regularization scheme is expected to change the details of the wave function near $r \sim \mathcal{R}$. But this does not invalidate the qualitative statement I made in Eq.~\eqref{eqn_V2pisuper} regarding the divergences of $\langle \psi_k | V_{2\pi} | \psi_k \rangle$.


\end{document}